# Hierarchichal-segmented AO in order to attain wide field compensation in the visible on an 8m class telescope

Roberto Ragazzoni[a,c], Demetrio Magrin[a,c], Jacopo Farinato[a,c], Marco Dima[a,c],
Davide Greggio[a,c], Renato Falomo[a], Valentina Viotto[a,c], Maria Bergomi[a,c],
Federico Biondi[a,c], Simonetta Chinellato[a,c], Marco Gullieuszik[a],
Luca Marafatto[a,c], Daniele Vassallo[a,b,c]

[a] INAF - Osservatorio Astronomico di Padova, Vicolo dell'Osservatorio 5, 35122 Padova, Italy
[b] Dipartimento di Fisica ed Astronomia - Università degli Studi di Padova,
Vicolo dell'Osservatorio 3, 35122 Padova, Italy
[c] ADONI – Laboratorio nazionale per l'Ottica Adattiva, Italy

## ABSTRACT

We describe the preliminary optimized layout for a partially optimized concept of an optical-8m class VLT-like 2x2 segmented camera where each channel is assisted by an equivalent of an MCAO system where the ground layer correction is commonly employed while the high altitude ones is performed in an open-loop fashion. While we derive the basic relationships among the Field of View and attainable correction with a pre-defined choice for the hardware, we discuss sky coverage and wavefront sensing issues employing natural and artificial references, involving the latest state-of-the-art in the development of wavefront sensing. We show that a flexible approach allow for a compensated Field of View that is variable and can be properly tuned matching the current turbulence situation and the requirement in term of quality of the compensation. A preliminary description of the overall optomechanical package is given as well along with a rough estimates of the efforts required to translates such a concept into reality.

## 1. INTRODUCTION

In order to achieve ground-based diffraction limited imaging in nowadays large telescopes the technique of the Multi Conjugated Adaptive Optics (MCAO)[1,2] has been introduced almost three decades ago. Such an approach has been the subject of detailed investigation concerning their ability to actually compensate a certain Field of View under well defined conditions of observations (like the used wavelength and the target to be reached in terms of performances, as the Strehl ratio[3]) and seeing conditions. Furthermore, new techniques has been envisaged to actually achieve such a goal[4] eventually using the now ubiquitous pyramid wavefront sensor[5]. These concept eventually turned to actual devices looking at the sky with the state of the art telescopes[6,7] in the usual realm where Adaptive Optics achieve its best performance: the Near InfraRed. Extension to the visible bandwidth has been envisaged from long time (see for example[8]) also corroborated by the opening of scientific realms like astrometry, that usually were neglected to this kind of approach[9,10].

In this context, MCAO become enough mature to push its limits, on one hand toward the largest foreseen apertures, like in MAORY for the ELT[11] or to achieve it on existing 8m class telescopes but on a much shorter wavelength, opening in principle a straight competition with current and short term planned space telescopes facility like HST and JWST.

In the following we describe and examine a very preliminary concept for achieving such a goal on a VLT-like telescope. While the kind of science doable with an hypothetical visible, diffraction limited, wide field system is well beyond the limit and scopes of this manuscript, one can recall that a number of classical and forefront instrumentation, achieving for example spectroscopy on a grid of positions in the sky, would become available in the wavelength regime where most of the archive information in astronomy today exists.

## 2. A MATTER OF ACTUATORS

Most of the existing 8m class telescopes today has, or will have in the nar future, a built-in adaptive device, namely a secondary adaptive mirror[11]. This device is usually conjugated very close to the ground layer. In a conventional, single conjugated, AO system this conjugation height does not, at least at first order, really play a significant role in the process of image compensation but, in an MCAO context, it can only be devoted to the correction of most of the ground layer turbulence. In principle, such a ground correction, can be achieved in a way that the remaining of the MCAO system is left to compensate the residual turbulence, mostly distributed away from the ground, although including the detail of the compensation in the ground layer loop can take to some advantage.

In principle, by consequences, one can envisage an MCAO system with a certain number of Deformable Mirrors, DMs, conjugated to various height. Using the rule of thumb introduced by Beckers[1,2] one need as many DMs as how much time one want to increase the Field of View. Given that in the visible the isoplanatic patch is notoriously rather limited (of the order of ten arcsec, the actual figure heavily depending upon the seeing conditions and somehow from the turbulence profile) one is actually tempted to introduce a significant number of additional DMs. As these are optically conjugated to altitudes relatively rather close one to each other it is realistic that one would need a relay system per each of these. While this is just a technical matter, the system that could derive could become easily very awkward from a technical viewpoint, or, on the opposite, one would be tempted to limit the number of total DMs. Performances, from a purely AO viewpoint, depends only –to a certain extent- upon the overall number of actuators involved.

## 3. SEGMENTING THE FIELD OF VIEW

All the considerations carried out in the previous section leads to the concept, already explored in other contexts, to segment the Field of View, once it has been corrected for the ground layer disturbance by the Adaptive Secondary Mirror[12] properly driven in order to achieve the best correction for the full ensemble of the split Field of Views[13,14]. This is depicted in Fig.1 in the usual turbulence diagram with altitude and spatial frequency in the two axes.

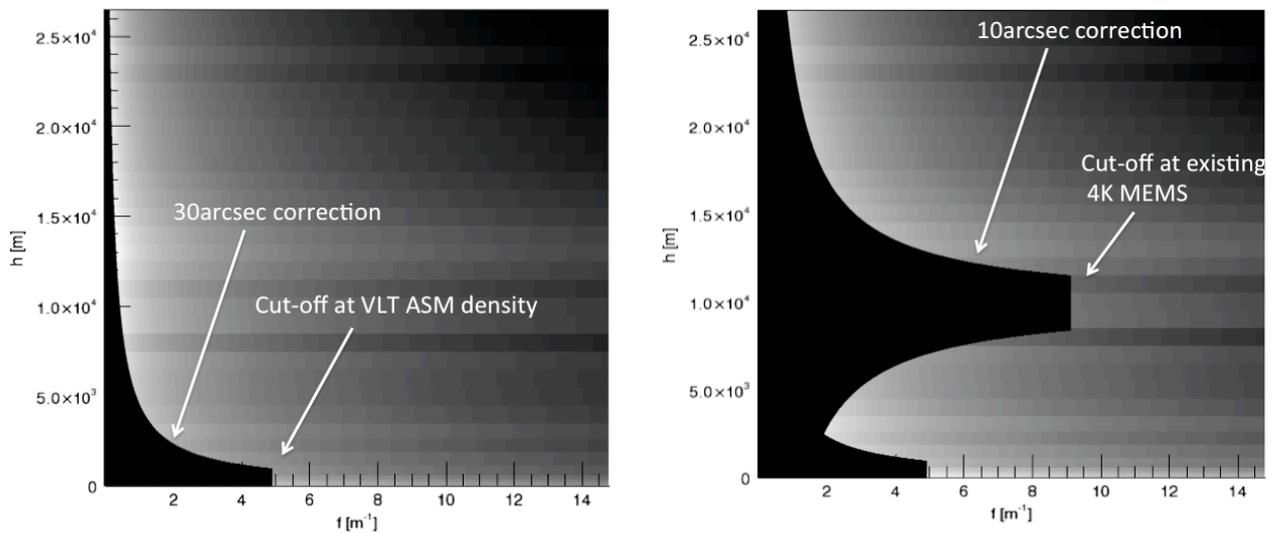

Figure 1: Left: a diagram showing the turbulence to be compensated, corrected by the Adaptive Secondary Mirror (ASM) using a ground layer correction to be effective within a full field of view of 30 arcsec and properly cut at the equivalent spatial sampling of the , relatively limited, density of the actuators of such a device. Right: On top of such a ground layer correction a large number of actuators DM using MEMS technology is compensating a much limited 10 arcsec wide Field of View, showed in the same turbulence diagram.

We initially carried out an hypothesis of segmenting the Field of View in 3 by 3 squared subapertures, each of a fixed size of the order of 10 arcsec. Wavefront sensing is achieved in a cooperative manner on both the artificial LGS and the (eventually in the Field of View) NGSs. The first are sensed using a reflective roof ingot-like approach[15] and we depicted in Fig.2 a possible layout for their arrangement. As the LGSs are monochromatic, their light is collected and folded by a dichroic and the WFSs are mounted on radially movable (to follow the different range of the LGSs because of the different altitude of the telescope) and tiltable (in order to match the equivalent optical length of the LGSs on the ingots). This is not the only option to achieve such a goal. One can equivalently use some trombones or an anamorphic system where one of the equivalent focal length along the axis aligned with the roof's edge can be tuned to match the equivalent LGSs length.

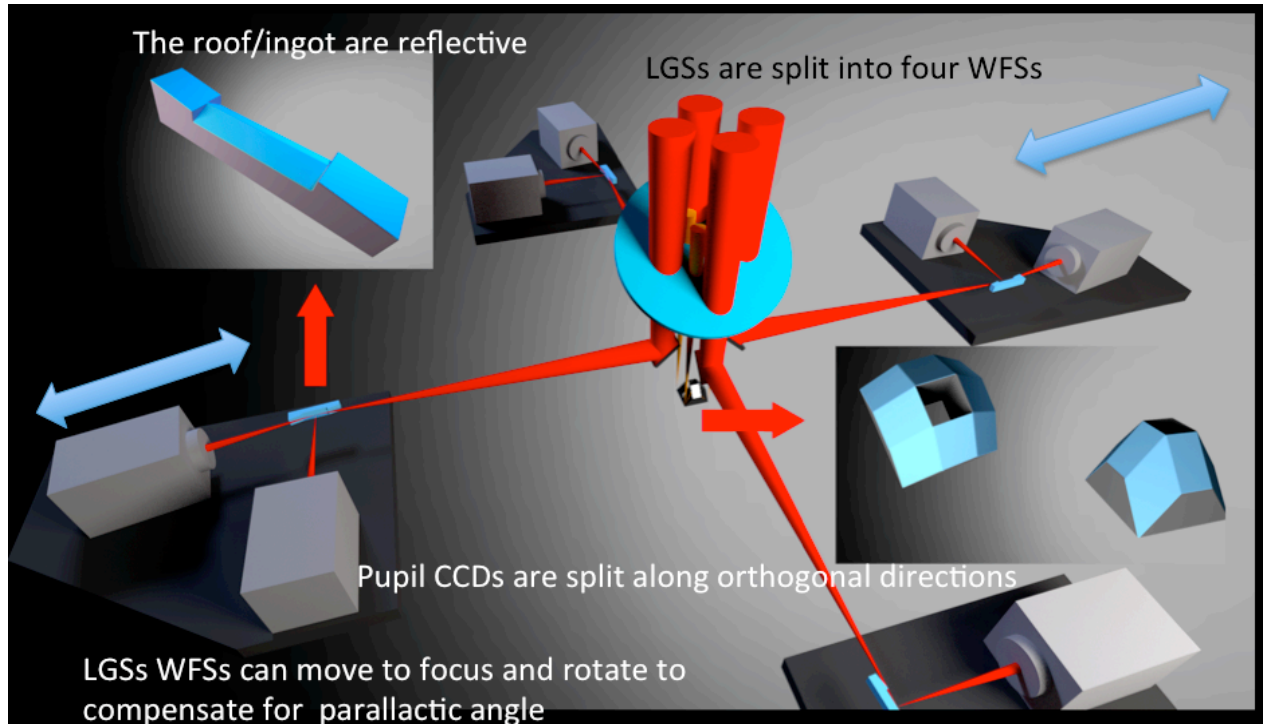

Figure 2 A large dichroic redirect the Na wavelength laser guide stars light onto an ensemble of WFSs of ingot-like choice. These can adjust for the different perspectives achieved at various elevations of the telescope with two movements with little loss in terms of efficiency.

In the explored concept we envisaged three layers that are described following the propagation of the light from the telescope onward the instrument.

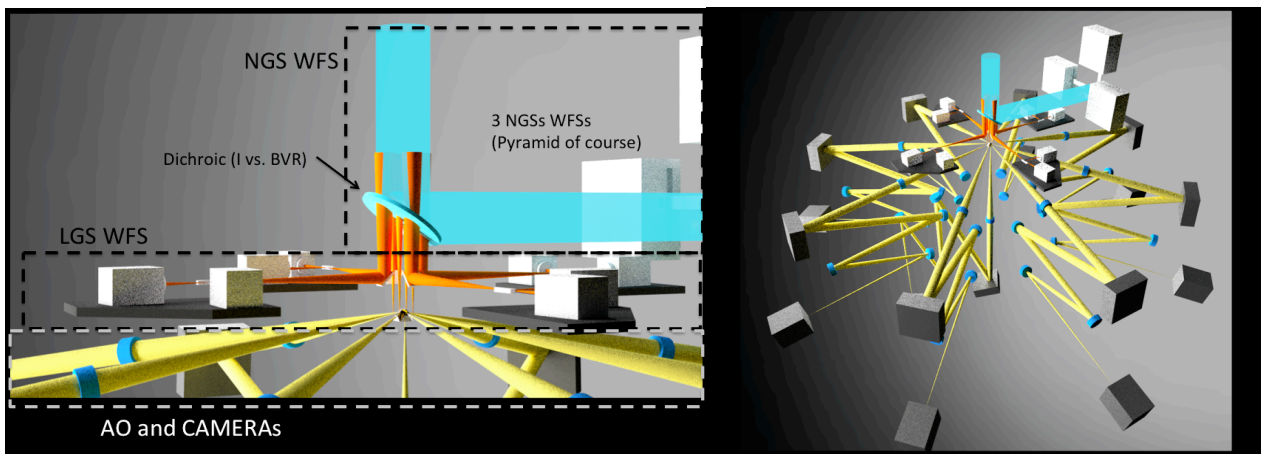

Figure 3 Left: a sectional view showing the two WFS layers, covring the NGS and the LGS, and the actual AO and science camera channels. Right: a 3D view of the whole instrument ensemble.

In the first one a set of pyramid WFSs analyze the light of NGSs that by chance are in the covered FoV. In the second layer the light is fed to the ingot-like WFSs analyzing the artificial Sodium reference light. We assumed, as per the VLT, to have four LGSs arranged in a squared shape around the line of sight of the telescope. In the third layer the Field of View is actually split by a small device consisting of a glass optical element with 8 facets mirrors and a central square hole. In this way there are 8 basically identically arms each evolving radially along the instrument. These comprises an optical relay, a 4K MEMS DM and an optical camera relay with at least one folding. The ninth central channel, is displaced along the optical axis of the whole ensemble and in fact it destroy a certain symmetry of the optomechanical layout. One can imagine to have all the nine modules to be mounted on a rack-like mechanical support so they can be serviced when any kind of failure would occurs. It is straightforward to think that a proper servicing would involve a tenth element that can be used to minimize the technical downtime of such device. Clearly the central arm makes the optomechanical choice for such serviceability not trivial. This concept is not further explored at this level, however.

We made a rough estimation of the overall amount of optomechanical elements in the approach described so far and it exceeds about 120 optical components. While this is just less than 10% of the overall optical components of others complex VLT instruments (like MUSE) it still represent a challenging figure. Furthermore, the subdivision and the size of the split Field of View is inherently rigid. We hence conceptually briefly investigated an option where a mere 2x2 splitting is being achieved. While this makes the overall number of arms (and related MEMS DMs) diminishing by more than a factor two, there is a crucial consequences: the covered Field of View can be made flexible by a combination of moving WFSs and adjusting the proper parameters in the tomographic correction to be achieved. In other words the outer edges of the four split Field of View are unbounded by any physical constraints (other than the practical implementation of the optical relay in the four second DM, MCAO arms) and depending upon the seeing conditions and/or the kind of science that one would like to achieve (requiring a different level of performances) one can define the whole Field of View (and by proper spatial filtering, the kind of commands to send to the ASM).

This approach is somehow sketched in Fig.4 along with a graphical representation, useful just for illustrative purposes without pretending to offer fixed numbers for reference, of such a scheme.

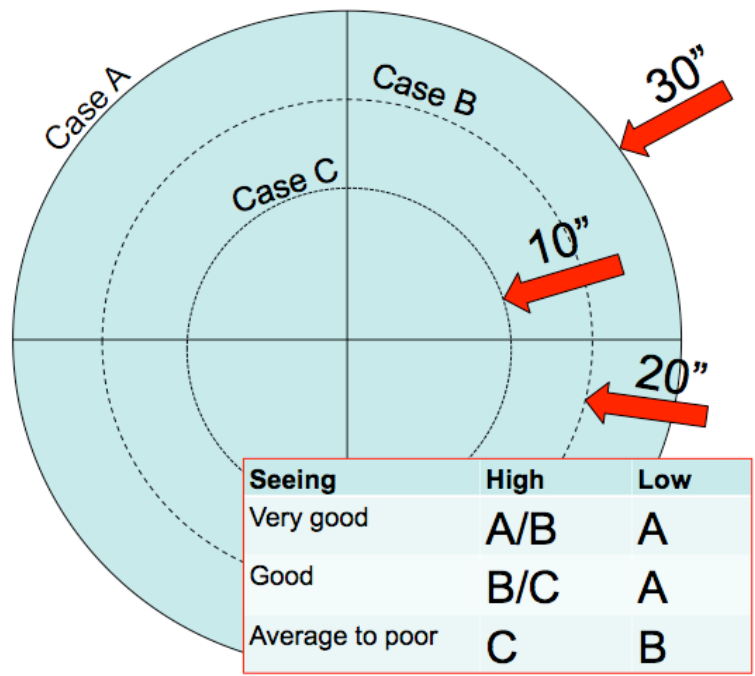

Figure 4: An illustrative example of how the splitting of the four quadrants (each being interested by an MCAO arm) can have their outer edge tuned to the conditions of the seeing and to the request (high or low) of the Strehl to be achieved to reach the scientific goal.

In principle each of the four arms could be equipped with two DMs rather than a single one, making the system more complex but with a better performances in terms of Strehl, although with a slightly limited throughput.

## 4. CONCLUSIONS

Pushing the limits to an MCAO system making it operational toward the visible is a task that is to be attacked by various sides. From the pure viewpoint of the performance achieved in terms of wavefront compensation there is obviously need of a large number of actuators. These have to match the Fried parameter that is particularly low for the visible regime, and because the equivalent isoplanatic patch is small, a good correction would requires a large number of DMs in an MCAO train or to accept a small corrected Field of View. Large Field of View using existing very tiny pitch DMs with large number of actuators (4K actuators today exists with MEMS technology) because of the Lagrange invariant requires very fast optics for pupil reimaging. Such optics, other than relatively complex, have a small optical depth, making hard to place contiguous DMs and very likely requiring per each DM in a conventional MCAO approach a relay. This will easily makes the overl throughput a miserable one, further to make alignment and operations rather complex. All this is wiped out with a segmented Field of View solution. Of course compromises are possible in any direction. We have shown that a 3x3 segmentation, although leading to a rather multiplexed instrument, sounds doable, while we envisage that a 2x2 solution, incorporating an inherent flexibility in the overall covered Field of View, has serious chance to be among the most practicable solutions for a visible wide field AO assisted instrument at an 8m class telescope like the VLT.